# Quantum-Probe Field Microscopy of Ultrafast Terahertz Excitations

Moritz B. Heindl[1], Nicholas Kirkwood[2], Tobias Lauster[3], Julia A. Lang[1],
Markus Retsch[3], Paul Mulvaney[2], and Georg Herink[1*]

[1] Experimental Physics VIII – Ultrafast Dynamics, University of Bayreuth, Germany
[2] ARC Centre of Excellence in Exciton Science, School of Chemistry, University of Melbourne, Australia
[3] Physical Chemistry I, University of Bayreuth, Germany

**Rapid evolutions of microscopic fields govern the majority of elementary excitations in condensed matter and drive microelectronic currents at increasing frequencies. Beyond nominal "radio frequencies", however, access to local electric waveforms remains a challenge. Several imaging schemes resolve sub-wavelength fields up to multi-Terahertz (THz) frequencies - including scanning-probe techniques, electro-optic sampling or recent ultrafast electron microscopy. Yet, various constraints on sample geometries, acquisition speed and maximum fields limit applications. Here, we introduce ubiquitous far-field microscopy of ultrafast local electric fields based on drop-cast quantum-dot probes. Our approach, termed Quantum-probe Field Microscopy (QFIM), combines fluorescence imaging of visible photons with phase-resolved sampling of electric fields deeply in the sub-wavelength regime. We capture stroboscopic movies of localized and propagating ultrafast Terahertz excitations with sub-picosecond temporal resolution. The scheme employs field-driven modulations of optical absorption in colloidal quantum-dots via the quantum-confined Stark-effect, accessible via far-field luminescence. The QFIM approach is compatible with strong-field sample excitation and sub-micrometer resolution – introducing a route towards ultrafast field imaging in active nanostructures during operation.**

Electro-magnetic waveforms hide from human perception and first proof of their existence required conversion to visible fluorescence by Heinrich Hertz[1]. Today, electric waveforms are coherently sampled by virtue of ultrashort laser pulses[2–4] – directly accessing the temporal signatures of swift charge motion and (quasi-)particle excitations in condensed matter systems up to the visible spectrum[5]. Yet, relevant field distributions are often confined to microscopic scales below the diffraction limit – arising from inhomogeneity of materials, microstructure design or intrinsic confinement of light-matter excitations[6–8]. Several approaches resolve local electric near-field waveforms up to multi-Terahertz frequencies, including raster-scanned photoconductive switches and electro-optic microscopy reaching micrometer scales[9–12]. Significantly enhanced resolution is provided by scanning near-field optical microscopy[13–15] (SNOM), THz-driven scanning tunneling microscopy[16,17] or recently emerging ultrafast electron microscopy[18–20], however, at the expense of significantly increased complexity.

Here, we demonstrate ultrafast optical far-field microscopy of electric fields up to THz frequencies. We image visible photons from local quantum-dot probes and acquire stroboscopic movies of electric near-field evolutions. The scheme employs the quantum-confined Stark effect[21–23], encoding electric near-fields via quasi-instantaneous variations of photo-absorption into far-field luminescence modulations, illustrated in Fig. 1. In essence, we present an ultrafast and coherent version of Hertz's experiments at 10.000-times higher frequency[1]. The feasibility of THz-induced coherent interactions was previously reported for diverse 0D-quantum systems[24–26]. Harnessing this mechanism, we perform spatially-resolved time-domain spectroscopy, and demonstrate the ultrafast imaging capabilities by resolving i) the localized THz-excitation of a bowtie antenna, ii) the transient mode pattern of a split-ring resonator and iii) propagating THz-excitations inside micro-slit waveguides. The experiments are based on two-color excitation using single-cycle Terahertz pulses for driving phase-stable near-fields in microstructures and visible fs-pulses for resonant absorption in the quantum dot probes, see Fig. 1a. The driving THz pulses at incident electric field strengths up to 400 kV/cm are resonantly enhanced in lithographically patterned gold antennas. Colloidal CdSe-CdS core-shell nanocrystals are deposited as quantum-probes analogous to low-frequency voltage sensing applications[27,28]. Luminescence is excited via wide-field illumination in the image plane of a fluorescence microscope with ~150 fs pulses at wavelengths around 500 nm. Differential images of the emission yield are acquired with a CCD camera in presence and absence of THz-excitation. The difference signal represents the key observable for instant local fields, which we refer to as the QFIM signal $S_{QFIM}$ in the following.

*Corresponding author: Georg.Herink@uni-bayreuth.de

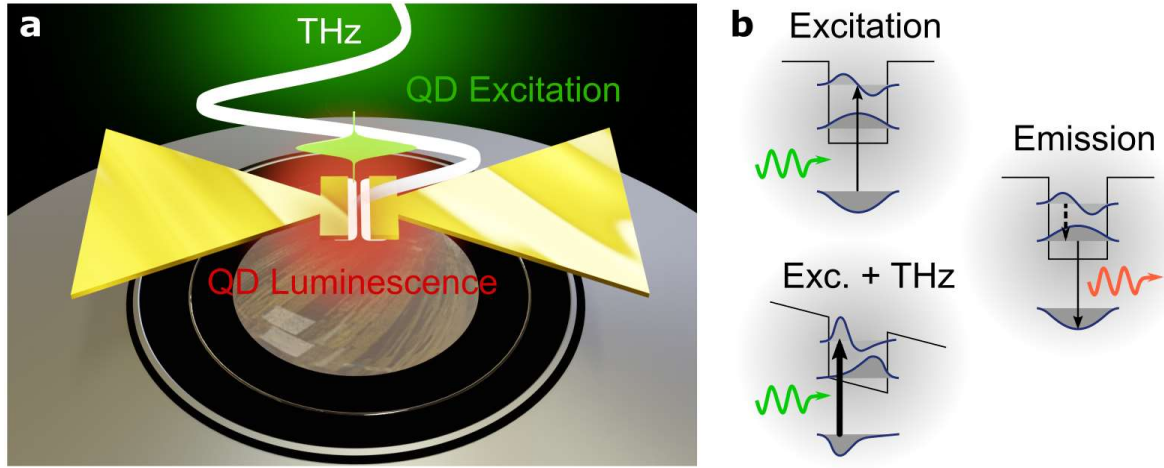

**Figure 1 Quantum-Probe Field Microscopy (QFIM): a)** THz electric near-fields are imaged in a fluorescence microscope using quantum-dot luminescence. Ultrashort sampling pulses experience modulated absorption due to the quantum-confined Stark effect, as depicted in **b)**. For example, increased absorption in the presence of local THz fields translate to higher luminescence rates, accessible by optical microscopy.

Now, we follow ultrafast near-field evolutions inside an antenna structure with sub-cycle temporal resolution by acquiring a sequence of snapshot images at increasing delays between THz and visible pulses. Figure 2a shows nine exemplary frames out of a series with temporal separation of $\Delta \tau = 30 fs$. We observe a strong enhancement in the antenna gap and in proximity to both terminal bars at THz polarization set ~0° to the antenna axis. The signal is maximized at the edge of each antenna leg and decays symmetrically towards the center of the bowtie. The detected pattern visually matches finite-element simulations of the THz near-field, shown in Fig. 2b, and characteristically depends on incident polarization (data for sample rotation of 90° in Supplementary Information). Based on our simulations, we estimate a maximum effective intra-gap near-field strength of ~10 MV/cm.

Analyzing the QFIM signal inside the gap, we demonstrate the extraction of local electric waveforms and characterize the temporal response of the bowtie antenna. As a prerequisite, we study the connection between field strength $F$ and $S_{QFIM}$: Measurements with varying incident field strengths yield the dependence $S_{QFIM}(F) \propto F^{1.9}$, as evident in the double-logarithmic representation in Fig. 3b. Employing this rectifying relation and the incident far-field waveform – obtained from conventional EO-sampling – , we are able to characterize the electro-magnetic transfer function of the structure[29–31]. In particular, we find that a resonance at 1.3 THz, see Fig. 3a, induces a local waveform that closely reproduces detailed features of the experimental QFIM trace. Finally, the comparison of the incident THz waveform and the extracted local field evolution is shown in Fig. 3c.

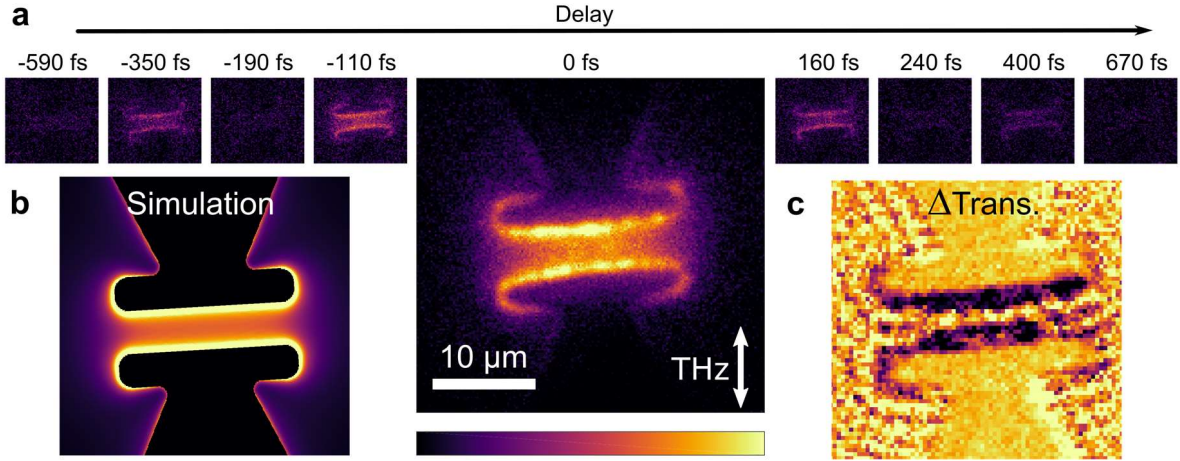

**Figure 2 Evolution of local fields at a resonant bowtie antenna: a)** Series of subsequent microscopic snapshots at selected delays, tracking three THz near-field cycles. Incident THz polarization as indicated. **b)** Simulated spatial near-field distribution at resonance in the gap region closely resembling the QFIM signal. **c)** A transient transmission snapshot (differential rel. signal), taken at the peak field $\Delta\tau = 0\ fs$, directly relates the QFIM signal to the field-driven enhancement of QD-absorption.

The underlying mechanism enabling the QFIM scheme relies on THz-driven modulations of the electronic band structure in low-dimensional quantum systems[21,22], i.e. the quantum-confined Stark effect in semiconductor nanocrystals[23].The altered electron and hole wavefunctions induce a quasi-instantaneous change of the transition dipole moment. Depending on the visible excitation frequency and accessed electronic state, absorption is either enhanced or reduced, as previously resolved via transient absorption spectroscopy[26]. We image these changes via luminescence emission microscopy. Specifically, we note that irrespective of much longer luminescence lifetimes around 10 ns, the temporal sampling resolution is exclusively governed by the ultrafast excitation process. Moreover, we verify this quasi-instantaneous nature of the absorption mechanism via imaging of the antenna in transmission mode at the peak of the local THz waveform in Fig. 2c, reproducing the complementary QFIM emission pattern.

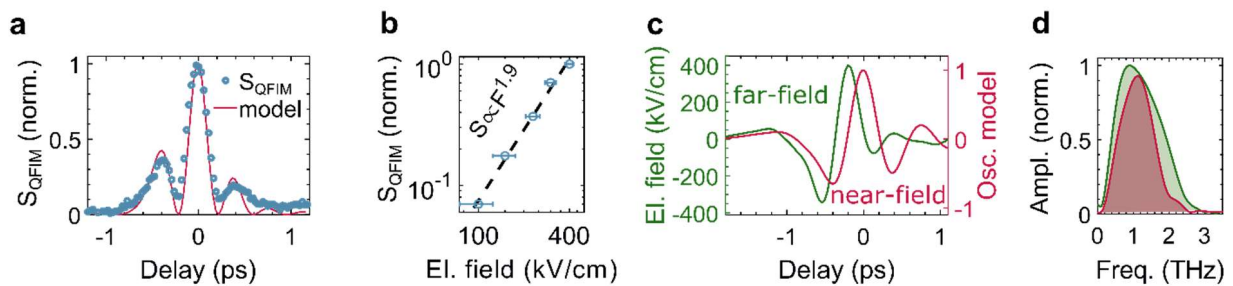

**Figure 3 Extraction of local waveforms: a)** Local QFIM signal in the gap of the bowtie (blue circles) and the modeled temporal luminescence evolution (red). **b)** Characterization of the QFIM signal (circles) as a function of the incident field strength. **c)** Comparison of incident temporal waveform (green) and extracted local field (red). **d)** Respective Fourier spectra of the waveforms in c).

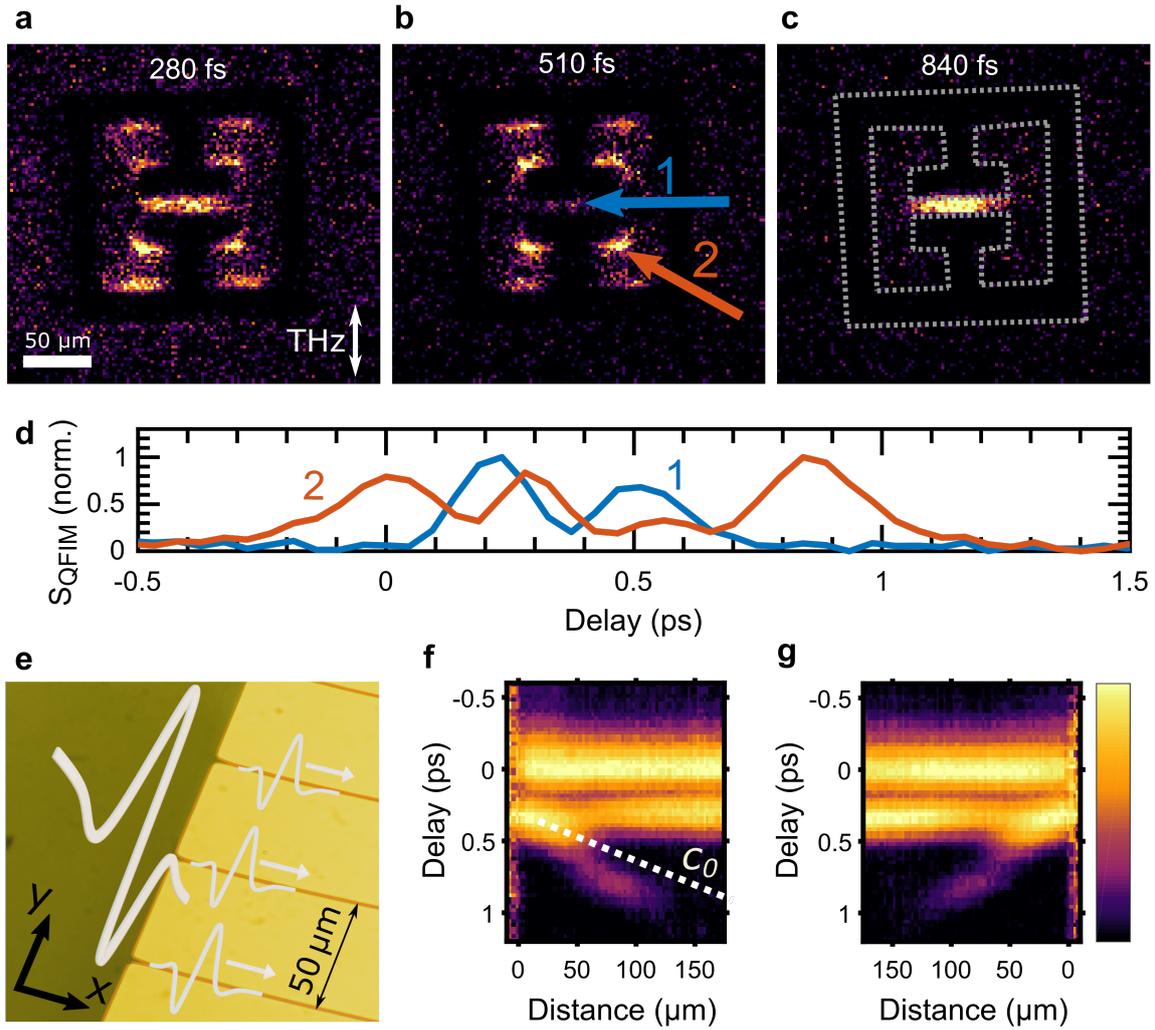

**Figure 4 Mapping spatially evolving near-fields: a)-c)** Consecutive QFIM snapshots of a THz-driven split-ring resonator, revealing two patterns with distinct temporal structure. **d)** Extracted local QFIM waveforms of the regions indicated in b). **e)** Edge-excitation of THz microslit waveguides. **f)** 2D representation of the normalized QFIM waveforms obtained by spatially integrating along the y-coordinate and concatenating subsequent delays. We track a propagating excitation along the waveguide with a velocity close to the vacuum speed of light indicated by the white dashed line. **g)** Symmetric propagation for excitation at the opposite side of the waveguide.

Next, we apply QFIM to track more complex spatial evolutions of electric near-fields in two distinct structures excited by single-cycle THz pulses. First, we resolve the near-field in an electric split-ring resonator by capturing QFIM snapshots, shown for three exemplary delays in Figs. 4a-c. We observe a characteristic near-field distribution with strong enhancement inside the ring structure and specifically in the resonator gap (see Fig. 4a). The full temporal dataset, however, reveals the independent evolution of these two features, as evident i.e. at instants $\Delta\tau = 510\ fs$ and $\Delta\tau = 840\ fs$ in Figs. 4b,c. The integrated QFIM signals of these regions are extracted and presented in Fig. 4d, resolving two individual temporal responses. These evolutions can be attributed to the coupling of a LC- and a dipole mode, a characteristic feature of this resonator geometry[32].

Finally, we study the pulsed excitation of a THz-waveguide, implemented via microgaps of ~2 μm width between gold films as shown in Fig. 4e. The THz focus is placed at the left edge of the structure. We display the corresponding QFIM signal in Fig. 4f in a combined spatio-temporal representation: the horizontal and vertical axis indicate the distance from the edge and the temporal delay, respectively. The dataset resolves two separate features: Initially, the structure shows a homogeneous response in the gaps extending over the THz-focus. Subsequently, a delayed feature emerges from the edge of the structure and appears at linearly increasing delay as a function of the distance. Analogously, excitation at the right edge of the waveguide yields the mirror symmetric dataset in Fig. 4g compared to f. The features correspond to propagating excitations with a velocity close to the speed of light ($c_0$, white dashed line) launched at the edges. This observation motivates future applications of QFIM for imaging complex radiation-matter couplings and propagating THz-excitations such as surface polaritons[7].

In conclusion, we introduce quantum-probe electric field microscopy and resolve ultrafast near-field evolutions up to 3 THz. Our approach combines the encoding of momentary THz-fields onto visible luminescence of nanocrystals and far-field fluorescence imaging. We explicitly verify the field-driven nature of the employed luminescence observable. On this basis, we demonstrate the mapping of evolving near-field distributions in the gap of a bowtie antenna deeply in the sub-wavelength regime, local multi-mode patterns of a split-ring resonator and the propagation of THz-excitations along micro-slit waveguides. This versatile scheme is readily applicable for tracking ultrahigh-frequency electronic device operation, resolving nearfield-driven nonlinear dynamics and propagating light-matter excitations.

## Acknowledgments

We thank J. Köhler and M. Lippitz for experimental equipment and valuable discussions. This work was funded by the Deutsche Forschungsgemeinschaft (DFG, German Research Foundation), via project 4037115411.

## References

1. Hertz, H. Ueber die Ausbreitungsgeschwindigkeit der electrodynamischen Wirkungen. *Ann. Phys.* **270**, 551–569 (1888).
2. Wu, Q. & Zhang, X. -C. Free-space electro-optic sampling of terahertz beams. *Appl. Phys. Lett.* **67**, 3523–3525 (1995).
3. Keiber, S. *et al.* Electro-optic sampling of near-infrared waveforms. *Nat. Photonics* **10**, 159–162 (2016).
4. Leitenstorfer, A., Hunsche, S., Shah, J., Nuss, M. C. & Knox, W. H. Detectors and sources for ultrabroadband electro-optic sampling: Experiment and theory. *Appl. Phys. Lett.* **74**, 1516–1518 (1999).
5. Goulielmakis, E. *et al.* Direct Measurement of Light Waves. *Science* **305**, 1267–1269 (2004).
6. Huber, A. J., Keilmann, F., Wittborn, J., Aizpurua, J. & Hillenbrand, R. Terahertz Near-Field Nanoscopy of Mobile Carriers in Single Semiconductor Nanodevices. *Nano Lett.* **8**, 3766–3770 (2008).
7. Basov, D. N., Fogler, M. M. & Abajo, F. J. G. de. Polaritons in van der Waals materials. *Science* **354**, (2016).
8. Lundeberg, M. B. *et al.* Tuning quantum nonlocal effects in graphene plasmonics. *Science* **357**, 187–191 (2017).
9. Hunsche, S., Koch, M., Brener, I. & Nuss, M. C. THz near-field imaging. *Opt. Commun.* **150**, 22–26 (1998).
10. Wächter, M., Nagel, M. & Kurz, H. Metallic slit waveguide for dispersion-free low-loss terahertz signal transmission. *Appl. Phys. Lett.* **90**, 061111 (2007).

11. Jepsen, P. U., Jacobsen, R. H. & Keiding, S. R. Generation and detection of terahertz pulses from biased semiconductor antennas. *JOSA B* **13**, 2424–2436 (1996).
12. Blanchard, F. *et al.* Real-time terahertz near-field microscope. *Opt. Express* **19**, 8277 (2011).
13. Chen, H.-T., Kersting, R. & Cho, G. C. Terahertz imaging with nanometer resolution. *Appl. Phys. Lett.* **83**, 3009–3011 (2003).
14. Ribbeck, H.-G. von *et al.* Spectroscopic THz near-field microscope. *Opt. Express* **16**, 3430–3438 (2008).
15. Eisele, M. *et al.* Ultrafast multi-terahertz nano-spectroscopy with sub-cycle temporal resolution. *Nat. Photonics* **8**, 841–845 (2014).
16. Cocker, T. L. *et al.* An ultrafast terahertz scanning tunnelling microscope. *Nat. Photonics* **7**, 620–625 (2013).
17. Peller, D. *et al.* Quantitative sampling of atomic-scale electromagnetic waveforms. *Nat. Photonics* **15**, 143–147 (2021).
18. Zewail, A. H. Four-Dimensional Electron Microscopy. *Science* **328**, 187–193 (2010).
19. Ryabov, A. & Baum, P. Electron microscopy of electromagnetic waveforms. *Science* **353**, 374–377 (2016).
20. Feist, A. *et al.* Ultrafast transmission electron microscopy using a laser-driven field emitter: Femtosecond resolution with a high coherence electron beam. *Ultramicroscopy* **176**, 63–73 (2017).
21. Mendez, E. E. *et al.* Effect of an electric field on the luminescence of GaAs quantum wells. *Phys. Rev. B* **26**, 7101–7104 (1982).
22. Miller, D. A. B. *et al.* Band-Edge Electroabsorption in Quantum Well Structures: The Quantum-Confined Stark Effect. *Phys. Rev. Lett.* **53**, 2173–2176 (1984).
23. Empedocles, S. A. & Bawendi, M. G. Quantum-Confined Stark Effect in Single CdSe Nanocrystallite Quantum Dots. *Science* **278**, 2114–2117 (1997).
24. Hoffmann, M. C., Monozon, B. S., Livshits, D., Rafailov, E. U. & Turchinovich, D. Terahertz electro-absorption effect enabling femtosecond all-optical switching in semiconductor quantum dots. *Appl. Phys. Lett.* **97**, 231108 (2010).
25. Pein, B. C. *et al.* Terahertz-Driven Luminescence and Colossal Stark Effect in CdSe–CdS Colloidal Quantum Dots. *Nano Lett.* **17**, 5375–5380 (2017).
26. Pein, B. C. *et al.* Terahertz-Driven Stark Spectroscopy of CdSe and CdSe–CdS Core–Shell Quantum Dots. *Nano Lett.* **19**, 8125–8131 (2019).
27. Park, K., Deutsch, Z., Li, J. J., Oron, D. & Weiss, S. Single Molecule Quantum-Confined Stark Effect Measurements of Semiconductor Nanoparticles at Room Temperature. *ACS Nano* **6**, 10013–10023 (2012).
28. Kuo, Y. *et al.* Characterizing the Quantum-Confined Stark Effect in Semiconductor Quantum Dots and Nanorods for Single-Molecule Electrophysiology. *ACS Photonics* **5**, 4788–4800 (2018).
29. Wang, K., Mittleman, D. M., van der Valk, N. C. J. & Planken, P. C. M. Antenna effects in terahertz apertureless near-field optical microscopy. *Appl. Phys. Lett.* **85**, 2715–2717 (2004).
30. Werley, C. A. *et al.* Time-resolved imaging of near-fields in THz antennas and direct quantitative measurement of field enhancements. *Opt. Express* **20**, 8551 (2012).
31. Wimmer, L. *et al.* Terahertz control of nanotip photoemission. *Nat. Phys.* **10**, 432–436 (2014).
32. Azad, A. K., Taylor, A. J., Smirnova, E. & O'Hara, J. F. Characterization and analysis of terahertz metamaterials based on rectangular split-ring resonators. *Appl. Phys. Lett.* **92**, 011119 (2008).